\documentclass{article}
\usepackage{amsmath,amsthm}
\usepackage{graphicx,epstopdf,epsfig,multirow,epic,bm}

\normalsize \rm
\begin{document}

\begin{center}
{\Large  \textbf {Trapping problem on star-type graphs with applications}}\\[12pt]
{\large Fei Ma$^{a,}$\footnote{~The author's E-mail: mafei123987@163.com. } \quad  and  \quad  Ping Wang$^{b,c,d,}$\footnote{~The author's E-mail: pwang@pku.edu.cn.} }\\[6pt]
{\footnotesize $^{a}$ School of Electronics Engineering and Computer Science, Peking University, Beijing 100871, China\\
$^{b}$ National Engineering Research Center for Software Engineering, Peking University, Beijing, China\\
$^{c}$ School of Software and Microelectronics, Peking University, Beijing  102600, China\\
$^{d}$ Key Laboratory of High Confidence Software Technologies (PKU), Ministry of Education, Beijing, China}\\[12pt]
\end{center}

\begin{quote}
\textbf{Abstract:} The trapping problem on graph (or network) as a typical focus of great interest has attracted more attention from various science fields, including applied mathematics and theoretical computer science, in the past. Here, we first study this problem on an arbitrary graph and obtain the closed-form formula for calculating the theoretical lower bound of average trapping time ($ATT$), a quantity that evaluates trapping efficiency of graph in question, using methods from spectral graph theory. The results show that the choice of the trap's location has a significant influence on determining parameter $ATT$. As a result, we consider the problem on star-type graphs, a special graph family which will be introduced shortly, with a single trap $\theta$ and then derive using probability generating functions the exact solution to quantity $ATT$. Our results suggest that all star-type graphs have most optimal trapping efficiency by achieving the corresponding theoretical lower bounds of $ATT$. More importantly, we further find that a given graph is most optimal only if its underlying structure is star-type when considering the trapping problem. At meantime, we also provide the upper bounds for $ATT$ of several graphs in terms of well-known Holder inequality, some of which are sharp. By using all the consequences obtained, one may be able to design better control scheme for complex networks from respect of trapping efficiency, to some extent, which are in well agreement with many other previous thoughts.

\textbf{Keywords:} Star-type graphs, Trapping problem, Average trapping time, Optimal trapping efficiency, Scale-free networks. \\

\end{quote}

\vskip 1cm

\section{Introduction}

The various dynamics on many complex systems, both natural and artificial, can be well described as random walks on an abstract object, known as complex network, including the stochastic behaviour of molecules in rarified gases \cite{Dongari-2011}, protein folding and misfolding \cite{Dobson-2003}, information flow in social networks \cite{Guille-2012}, traffic and mobility patterns on the internet \cite{Gonzalez-2008}, fluctuations in stock prices \cite{Wang-2010} as well as the behaviour of stochastic search algorithms \cite{Dorigo-2005},  and so forth. Hence, in the past several decades, random walks on complex networks have attracted increasing interest in science fields including applied mathematics, theoretical computer sciences, and statistical physics \cite{Li-2016}-\cite{Bertasius-2017}. Almost all previous works focus on an issues of how long it takes a random walker to reach a given target in the process of random walk. To address this problem, the quantity, called first-passage time ($FPT$), has defined in the literature and triggers a growing number of theoretical investigations over the past. It is using such a parameter that leads to important understanding of this paradigmatic dynamical process and uncovering the effects of the underlying network structure on the behavior of random-walk dynamics \cite{Volchenkov-2011}-\cite{Guerin-2016}.

Along the similar research, Montroll first introduced the trapping problem in his seminal work \cite{Montroll-1969} (defined in detail later). In fact, it is a kind of random walk where a perfect trap is allocated at a given position, absorbing all other particles (walkers) visiting it. As such, the trapping process is also closely associated with various other dynamical processes in diverse complex systems. Specifically, the trapping problem has been considered on regular lattices with different dimensions \cite{Garza-Lopez-2005}, the Sierpinski fractals \cite{Bentz-2010}, the T-graph \cite{Agliari-2008}, Cayley trees \cite{Wu-2012}, and Vicsek fractals as models of polymer networks \cite{Gurtovenko-2005}, to name but a few. The empirical observations and theoretical analysis both suggest that this problem may yield different results over networks with different characteristics, such as geometries, dimensions and time regimes \cite{Redner-2001}. More generally, the trapping problem on networks has been studied by Kittas \emph{et al} in Ref \cite{Kittas-2008}. Similar to the researches above, here, they calculated the evolution of the particle (walker) density $\rho(t)$ of random walkers in the presence of one or multiple traps with concentration $c$ and obtained some remarkable results, one of which is that in typical ER networks, while for short times $\rho(t)\propto\text{exp}(-Act)$ where $A=O(1)$, for longer times $\rho(t)$ exhibits a more complex behavior strong dependence on both the number of traps and the vertex number of network. In a nutshell, the underlying structure has a significant influence on determining the solutions to some parameters for describing the trapping problem on networks under consideration.

As known, a great number of real-world networks are proven to show scale-free feature \cite{Albert-1999-1} and small-world property \cite{Watts-1998}. As a result, the trapping problem has been considered by many researchers using some previous scale-free networked models \cite{Rieger-2004,Kittas-2008,Xing-2017,Zhang-2013}. The results reveal that for some correlated scale-free networks with a constant average degree $\langle k\rangle$ and a power-law degree distribution $P(k)\sim k^{-\gamma}$, the scaling of the lower bound of average trapping time $ATT$ (defined in section 2) to a hub vertex behaves
sublinearly with vertex number. At the same time, in scale-free networks, the particle (walker) density $\rho(t)\propto\text{exp}\left(-At/|V|^{\frac{\gamma-2}{\gamma-1}}\langle k\rangle\right)$ in which $|V|$ is the vertex number on network in question (defined in detail later).

Based on the results above, the goal of this paper aims at investigating the trapping problem on star-type graphs, which have wide applications in many areas, such as date center networks \cite{Couto-2016}. Additionally, many other interesting graphs are also chosen as our focuses. Note that, throughout this paper, the terms graph and network are used indistinctly for convenience.

The main contributions of this paper are as follows

(1) With methods from spectral graph theory, we instantiate that several types of specific graphs, including star graph $S_{n}$ and complete graph $K_{n}$, are able to show most optimal topological structure by satisfying a sharp theoretical lower bound for $ATT$, respectively.

(2) Based on the theoretical lower bound for $ATT$ on an arbitrary graph, we derive that the star-type graphs are able to show most optimal topological structure. And then, we further prove that an arbitrary graph $G(V,E)$ is most optimal with respect to the $ATT$ for trapping problem if and only if the underlying structure of graph $G(V,E)$ is star-type. As an immediate consequence, the exact solution to Kemeny's constant, another topological parameter that is close related to the trapping problem on graph in question, of the star-type graphs can be easily derived.

(3) With the help of results related to $ATT$ on star-type graphs, we obtain some exact solutions for $ATT$ on some graphs widely studied in the context of graph theory. Additionally, for practical applications as addressed in (4), we also provide the approximate solutions for $ATT$ on those graphs in terms of some typical techniques including Holder inequality. At meantime, some upper bounds for $ATT$ are not sharp because there exist no graphs such that the equality in approximate solution holds true.

(4) According to theoretical analysis mentioned above, we may be able to provide some effective measures for promoting the robustness of real-world networks by preventing a small number of vertices from outside attacks. On the one hand, these measures introduced here may have comparable performance to some previous ones used in context of complex network. On the other hand, one has the ability to find some shortcomings shown by a portion of pre-existing measures by means of the lights shed by our technical manners.

The rest of this paper can be organized by the next several sections. Section 2 introduces some basic terminologies, such as graph and its matrix representation, star-type graphs as well the trapping problem.  By using techniques from spectral graph theory,  we provide a precise deduction of the closed-from solution to the theoretical lower bound for average trapping time $ATT$ in Section 3. Meanwhile, star graph $S_{n}$ and complete graph $K_{n}$ are proved to have most optimal topological structure by achieving the corresponding theoretical lower bound for average trapping time $ATT$, respectively. To make further progress, in Section 4, we find the star-type graphs to display most optimal trapping efficiency via holding the theoretical lower bound for $ATT$. Also, we final conclude that all graphs but for star-type ones can not be most optimal with respect to the trapping problem under consideration. As a byproduct, here also reports the closed-form solution to Kemeny's constant for the star-type graphs. Next, the goal of Section 5 is to show some theoretical applications of scientific interest on the basis of results provided above, for instance, deriving the exact solutions to $ATT$ on some graphs of significant interest. In addition, some practical applications related to control schemes for the robustness of complex networks can be found in Section 6. Final, we close this paper in Section 7.

\section{Terminologies}

In this section, we will introduce some basic concepts and notations for graphs, star-type graphs, subdivision operation, as well the trapping problem on graphs. For convenience, it is worth noticing that throughout this paper, the terms graph and network are used indistinctly.

\begin{figure}
\centering
  \includegraphics[height=5cm]{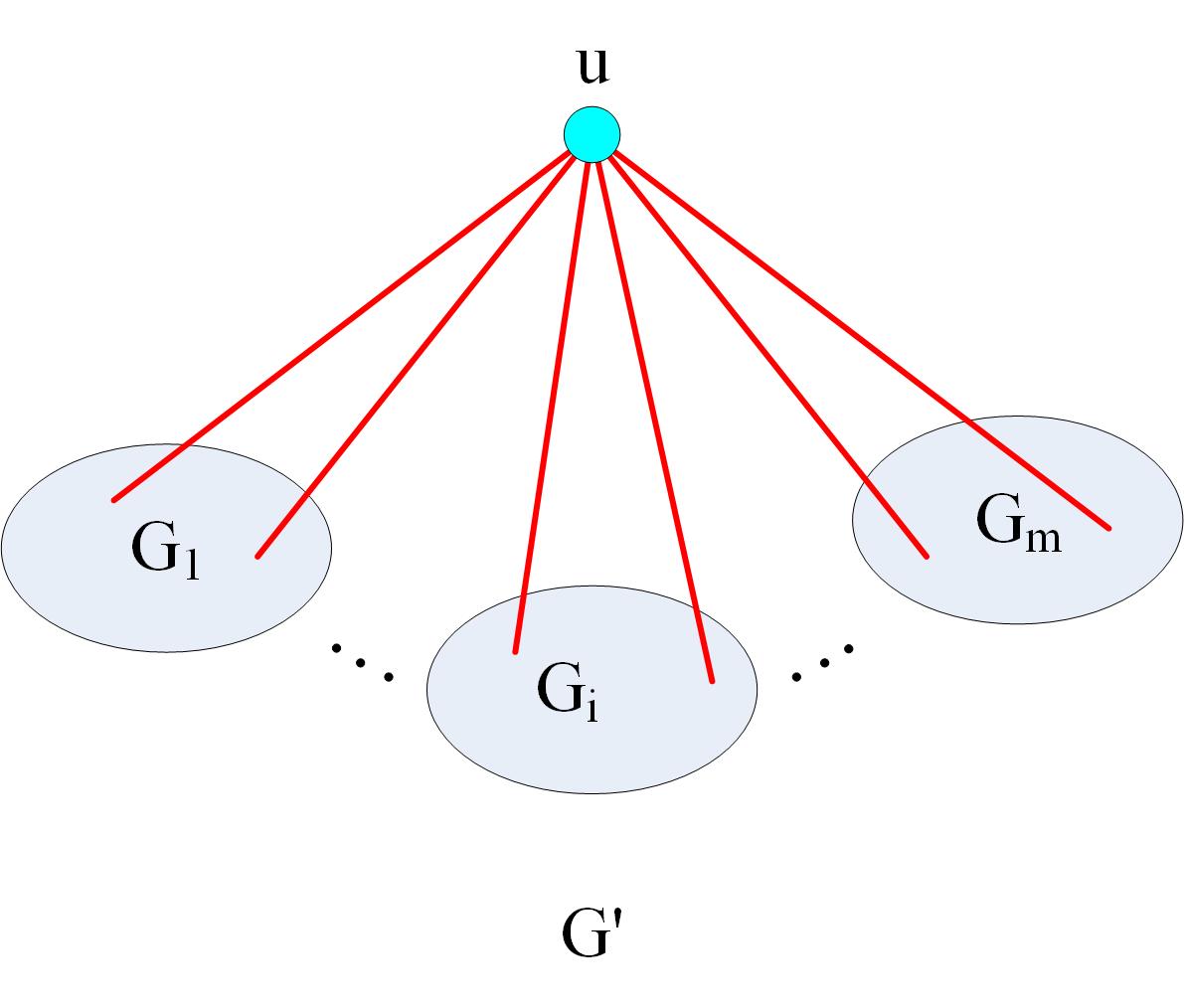}\\
{\small Fig.1. The diagram of Star-type graph $G'$.   }
\end{figure}

\subsection{Graph and its matrix representation}

A graph (or network) $G(V,E)$ is an ordered pair ($V(G),E(G)$) consisting of a set $V(G)$ of vertices and a set $E(G)$ of edges running between vertices. Unless otherwise specified, let $G$ denote a graph. The total number of vertices is denoted by $|V|$ and $|E|$ represents the edge number. Throughout this paper, all the discussed graphs are all simple and connected, namely, without multi-edges and loops.

More generally, it is convention to interpret a graph $G(V,E)$ by its adjacency matrix $\mathbf{A}=(a_{ij})$ in the following form

$$a_{ij}=\left\{\begin{aligned}&1, \quad\text{vertex $i$ is adjacent to $j$}\\
&0,\quad\text{otherwise}.
\end{aligned}\right.
$$
This thus contains some basic information about a graph itself, such as, the degree $d_{i}$ of vertex $i$ equal to $d_{i}=\sum_{j=1}^{|V|}a_{ij}$. Accordingly, the diagonal matrix, denoted by $\mathbf{D}$, may be timely defined as follows: the $i$th diagonal entry is $d_{i}$, while all non-diagonal elements are zero, i.e., $\mathbf{D}=\text{diag}[d_{1},d_{2},\dots,d_{|V|}]$.

\subsection{Star-type graph}

Given a set of graphs $G_{i}$ ($i=1,2,\dots,m$), bringing an external vertex $u$ and connecting it to each vertex $v_{i}$ in graph $G_{i}$ together produce a new graph $G'$ as plotted in Fig.1. The resulting graph $G'$ is defined as \emph{star-type graph}. Obviously, Star-type graph $G'$ will be a star graph $S_{m}$ when each component $G_{i}$ is an isolated vertex.

\subsection{Subdivision operation}

For an arbitrary graph $G(V,E)$, inserting a new vertex into each edge $e$ in edge set $E$ leads to a graph $G^{1}(V^{1},E^{1})$ that is the first-order subdivision one of graph $G(V,E)$. Such a procedure is often regarded as the \emph{first-order subdivision operation} for graph $G(V,E)$. Equivalently, the end graph $G^{1}(V^{1},E^{1})$ can be obtained from original graph $G(V,E)$ by replacing each edge $e_{uv}$ in $E$ by a path $p_{uwv}$ of length $2$ where $w$ is the newly inserted vertex. An illustrative example is shown in Fig.2.

Thus, a couple of equations associated with $|V^{1}|$ and $|E^{1}|$ can be given by

 \begin{equation}\label{eqa:MF-2-3-1}
|V^{1}|=|V|+|E|, \qquad |E^{1}|=2|E|.
\end{equation}

In general, inserting $n$ vertices into each edge $e$ in graph $G(V,E)$ results in a graph $G^{n}(V^{n},E^{n})$ that is called the \emph{n}th-order subdivision one of graph $G(V,E)$. Accordingly, this procedure is naturally viewed as the \emph{nth-order subdivision operation} for graph $G(V,E)$. As above, we have

 \begin{equation}\label{eqa:MF-2-3-2}
|V^{n}|=|V|+n|E|, \qquad |E^{n}|=(n+1)|E|.
\end{equation}

\begin{figure}
\centering
  \includegraphics[height=5cm]{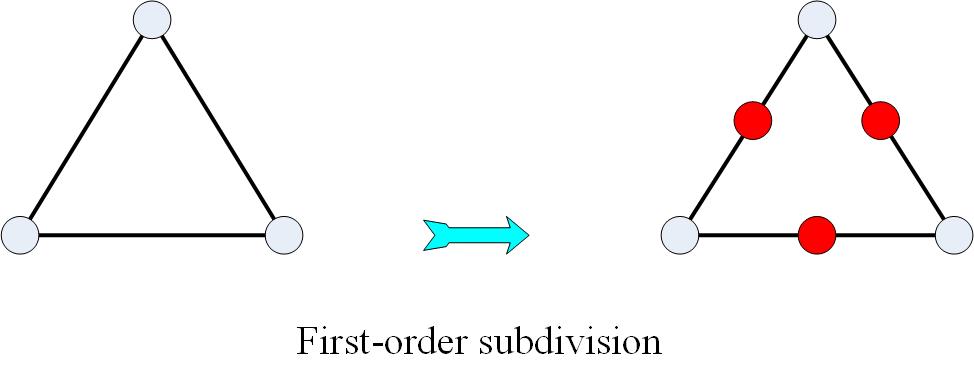}\\
{\small Fig.2. The diagram of first-order subdivision operation.   }
\end{figure}

\subsection{Trapping problem}

For a given graph $G(V,E)$, we may consider an unbiased discrete-time random walk taking placing on it. At each time step, a walker starting out from its current location $i$ moves with a uniform probability proportional to its degree $d_{i}$ to some vertex $j$ of its neighboring set \cite{Aldous-1999}. Specifically, such a stochastic process of random walks can be certainly represented by the transition matrix $\mathbf{P}=\mathbf{D}^{-1}\mathbf{A}$, whose entry $p_{ij}=a_{ij}/d_{i}$ indicates the probability of jumping from $i$ to $j$ in one step. The trapping problem is a specific case of random walks where an arbitrarily given vertex will be chosen as a deep trap $\theta$. For our purpose, let the term $\theta$ denote by the trap and its corresponding location vertex simultaneously. Generally speaking, the choice of location vertex is based on either practical applications or theoretical flavor, such as, assigning the trap on the greatest degree vertex.

Mathematically, in the trapping problem on graph $G(V,E)$ with a single trap $\theta$, a significant measure for a walker starting out from vertex $i$ is the trapping time $TT_{i\rightarrow\theta}$ that is in fact the expected time taken by the walker to first reach the trap $\theta$. As a result, for the whole graph $G(V,E)$ in question, the average trapping time $ATT_{\theta}$ can be defined as the averaged value over $TT_{i\rightarrow\theta}$ for all vertices $i$ and is written as

\begin{equation}\label{eqa:MF-2-4-1}
ATT_{\theta}=\frac{1}{|V|-1}\sum_{i\in V, i\neq \theta} TT_{i\rightarrow \theta}.
\end{equation}
This capsules a great deal of information about random-walk dynamics of this kind as will be shown shortly.

\section{The theoretical lower bound for $ATT$ and examples}

As previously, there is a long history for the study of trapping problem on graphs \cite{Wu-2012,Kittas-2008,Xing-2017,Burnashev-2012,Rhee-2011}. Indeed, there have been a larger number of interesting consequences associated with the trapping problem reported in the past. One of which is that a graph under consideration can be considered most optimal if its $ATT$ is precisely equal to the corresponding theoretical lower bound. Therefore, it is considerably significant to analytically deduce the expression for theoretical lower bound for $ATT$ on an arbitrary graph. As shown shortly, below proposes a detailed computation for the exact solution to theoretical lower bound for $ATT$ in a technical manner from spectral graph theory. At the same time, this section also shows that several classes of graphs, for example, star graph $S_{m}$ plotted in Fig.3, indeed exhibit most optimal trapping efficiency.

\subsection{Theoretical lower bound for $ATT$}

\textbf{Lemma 1} \emph{For an arbitrary graph $G(V,E)$ with a trap $\theta$, the theoretical lower bound for $ATT_{\theta}$ is given by}

 \begin{equation}\label{eqa:MF-3-1-1}
ATT_{\theta}\geq\frac{2|E|}{d_{\theta}}-1
\end{equation}
\emph{where $d_{\theta}$ is the degree of vertex occupied by that trap. It is worth noting that, hereafter, $d_{\theta}$ will be simply regarded as the degree of that trap for brevity.}

\emph{Proof } Given a graph $G(V,E)$ with a trap $\theta$, by using the techniques from spectral graph theory, if let $1=\lambda_{1}>\lambda_{2}\geq \lambda_{3}\geq \dots\geq \lambda_{|V|}$ be the $|V|$ eigenvalues of matrix
$\Gamma=D^{-\frac{1}{2}}AD^{\frac{1}{2}}$ and we then denote by $\psi_{1}, \psi_{2}, \psi_{3},\dots, \psi_{|V|}$ the corresponding mutually orthogonal eigenvectors of unit length in which $\psi_{i}=(\psi_{i1},\psi_{i2},\dots,\psi_{i|V|})^{\top}$. After that, the concrete expression of $ATT_{\theta}$ for graph $G(V,E)$  can be written as

\begin{equation}\label{eqa:MF-3-1-2}
\begin{aligned}ATT_{\theta}&=\frac{1}{1-\pi_{\theta}}\sum_{j=1}^{|V|}\pi_{j}2|E|\sum_{i=2}^{|V|}\frac{1}{1-\lambda_{i}}\left(\frac{\psi_{i\theta}^{2}}{d_{\theta}}-\frac{\psi_{ij}\psi_{i\theta}}{\sqrt{d_{j}d_{\theta}}}\right)\\
&=\frac{1}{1-\pi_{\theta}}\sum_{i=2}^{|V|}\left(\frac{1}{1-\lambda_{i}}\psi_{i\theta}^{2}\sum_{j=1}^{|V|}\frac{d_{j}}{d_{\theta}}\right)-
\frac{1}{1-\pi_{\theta}}\sum_{i=2}^{|V|}\left(\frac{1}{1-\lambda_{i}}\psi_{i\theta}\sqrt{\frac{2|E|}{d_{\theta}}}\sum_{j=1}^{|V|}\psi_{ij}\sqrt{\frac{d_{j}}{2|E|}}\right)
\end{aligned}
\end{equation}
where $\pi_{i}$ is equal to $d_{i}/\sum_{j=1}^{|V|}d_{j}$ and $2|E|$ equals $\sum_{j=1}^{|V|}d_{j}$. With the following fact

\begin{equation}\label{eqa:MF-3-1-3}
\sum_{k=1}^{|V|}\psi_{jk}\psi_{ik}=\sum_{k=1}^{|V|}\psi_{kj}\psi_{ki}=\left\{\begin{aligned}&1, \qquad \text{if $j=i$}\\
&0, \qquad \text{otherwise}\end{aligned}\right.,
\end{equation}
we can without difficulty find

\begin{equation}\label{eqa:MF-3-1-4}
\sum_{j=1}^{|V|}\psi_{ij}\sqrt{\frac{d_{j}}{2|E|}}=\sum_{j=1}^{|V|}\psi_{i j}\psi_{1j}=0.
\end{equation}
Here we have taken advantage of $\psi_{1}=(\psi_{11},\psi_{12},\dots,\psi_{1|V|})^{\top}=\left(\sqrt{\frac{d_{1}}{2|E|}},\sqrt{\frac{d_{2}}{2|E|}},\dots,\sqrt{\frac{d_{|V|}}{2|E|}}\right)^{\top}$. Therefore, Eq.(\ref{eqa:MF-3-1-2}) may be rearranged as

\begin{equation}\label{eqa:MF-3-1-5}
ATT_{\theta}=\frac{1}{1-\pi_{\theta}}\sum_{i=2}^{|V|}\left(\frac{1}{1-\lambda_{i}}\psi_{i\theta}^{2}\sum_{j=1}^{|V|}\frac{d_{j}}{d_{\theta}}\right).
\end{equation}

For more details about the deductions above to see Ref.\cite{Aldous-1999}. After that, using both some simple arithmetics and the well-known Cauthy's inequality,

\begin{equation}\label{eqa:MF-3-1-6}
\left(\sum_{i=2}^{|V|}\frac{1}{1-\lambda_{i}}\psi_{i\theta}^{2}\right)\left(\sum_{i=2}^{|V|}(1-\lambda_{i})\psi_{i\theta}^{2}\right)\geq \left(\sum_{i=2}^{|V|}\psi_{i\theta}^{2}\right)^{2},
\end{equation}
where $\sum_{i=2}^{|V|}(1-\lambda_{i})\psi_{i\theta}^{2}=1-\sum_{i=1}^{|V|}\lambda_{i}\psi_{i\theta}^{2}=1$ and $\sum_{i=2}^{|V|}\psi_{i\theta}^{2}=\sum_{i=1}^{|V|}\psi_{i\theta}^{2}-\pi_{\theta}=1-\pi_{\theta}$,
we can derive the theoretical lower bound for $ATT_{\theta}$ as below

\begin{equation}\label{eqa:MF-3-1-7}
ATT_{\theta}\geq\frac{1}{1-\pi_{\theta}}\frac{\sum_{j=1}^{|V|}d_{j}}{d_{\theta}}(1-\pi_{\theta})^{2}=\frac{\sum_{j=1}^{|V|}d_{j}}{d_{\theta}}-1.
\end{equation}
By far, this completes the proof of Lemma 1.

Roughly speaking, in the trapping problem considered here, the closer to the lower bound the value $ATT_{\theta}$ is, the more optimal the topological structure of the corresponding graph is. In essence, the theoretical lower bound for $ATT_{\theta}$ is sharp since there are many graphs holding the equality in the right-hand side of Eq.(\ref{eqa:MF-3-1-1}) as shown later.

\subsection{Several examples}

Here, we take several small graphs shown in Fig.3 as examples to illustrate concrete computation for average trapping time $ATT_{\theta}$ according to their own special structures.

\begin{figure}
\centering
  \includegraphics[height=4cm]{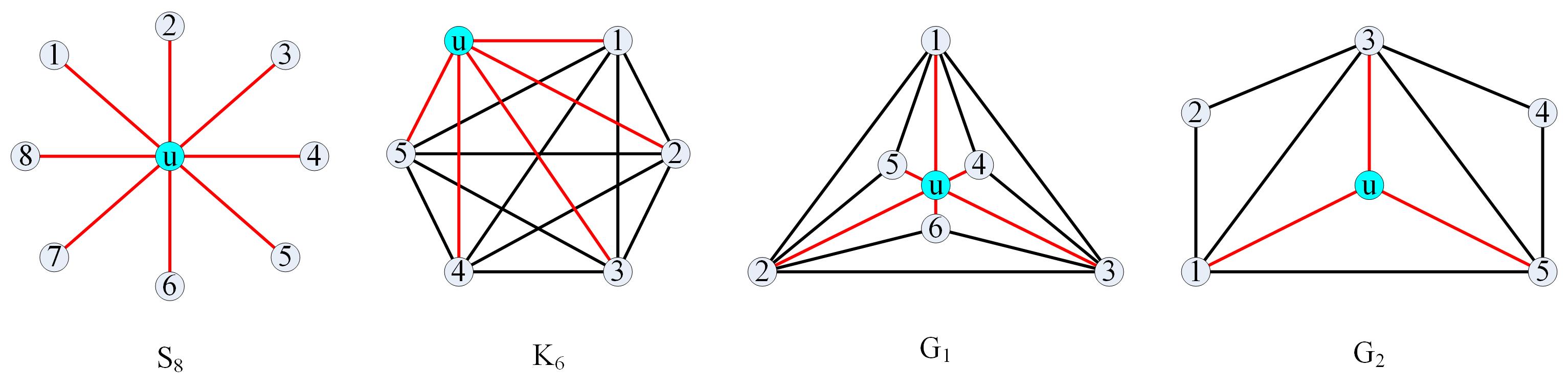}\\
{\small Fig.3. The diagram of several small graphs.   }
\end{figure}

\textbf{Example 1} \emph{For a star graph $S_{m}$ in which the trap $\theta$ is allocated on the central vertex $u$, the average trapping time $ATT^{(1)}_{\theta}$ is}

\begin{equation}\label{eqa:MF-3-2-1-1}
ATT^{(1)}_{\theta}=1.
\end{equation}

\emph{Proof } It is straightforward to see that a walker originating from a arbitrary leaf vertex $i$ of star graph $S_{m}$ must spend a step jumping to the center, i.e., $TT_{i\rightarrow \theta}\equiv 1$. By definition in Eq.(\ref{eqa:MF-2-4-1}), the $ATT^{(1)}_{\theta}$ is precisely given by
 \begin{equation}\label{eqa:MF-3-2-1-2}
ATT^{(1)}_{\theta}=\frac{1}{|V(S_{m})|-1}\sum_{i\in V, i\neq \theta} TT_{i}\equiv1.
\end{equation}

Clearly, star graph $S_{8}$ in the left-most panel of Fig.3 has average trapping time $ATT_{\theta}$ equal to $1$. It is worth noting that there are strong symmetrical characteristics observed on star graph $S_{m}$. In other words, each leaf vertex has completely the same both local and whole neighbor structure as all others. Similarly, we can find the exact solution to average trapping time $ATT^{(2)}_{\theta}$ on complete graph $K_{n}$ as below.

\textbf{Example 2} \emph{For a complete graph $K_{n}$ in which the trap $\theta$ is allocated on an arbitrary vertex, for instance, the vertex $u$ in the second panel of Fig.3, the average trapping time $ATT^{(2)}_{\theta}$ is
}
\begin{equation}\label{eqa:MF-3-2-2-1}
ATT^{(2)}_{\theta}=n-1.
\end{equation}

Here we omit detailed proof for brevity.

By analogy with the demonstrations above, one can timely obtain the closed-form expressions for the average trapping time on graphs $G_{1}$ and $G_{2}$ shown in Fig.3, respectively. The results are listed in the following examples.

\textbf{Example 3} \emph{For the graph $G_{1}$ in which the trap $\theta$ is allocated on vertex $u$, the average trapping time $ATT^{(3)}_{\theta}$ is}

\begin{equation}\label{eqa:MF-3-2-3-1}
ATT^{(3)}_{\theta}=4.
\end{equation}

\textbf{Example 4} \emph{For the graph $G_{2}$ in which the trap $\theta$ is allocated on vertex $u$, the average trapping time $ATT^{(4)}_{\theta}$ is}

\begin{equation}\label{eqa:MF-3-2-4-1}
ATT^{(4)}_{\theta}=\frac{67}{11}.
\end{equation}

We still intend to omit the concrete proofs for Eqs.(\ref{eqa:MF-3-2-3-1}) and (\ref{eqa:MF-3-2-4-1}) particularly because they can be easily obtained by hand.

Now, let us derive the formulas for theoretical lower bounds for average trapping time $ATT$ on graphs introduced in examples 1-4. For convenience, we denote by symbol $ATT'^{(i)}_{\theta}$ the theoretical lower bound for the value $ATT^{(i)}_{\theta}$. After that, the value for $ATT'^{(i)}_{\theta}$ can be immediately obtained upon both Eq.(\ref{eqa:MF-3-1-1}) and illustrations in Fig.3, as follows

\begin{equation}\label{eqa:MF-3-2-5-1}
ATT'^{(1)}_{\theta}=1, \quad ATT'^{(2)}_{\theta}=n-1, \quad ATT'^{(3)}_{\theta}=4, \quad \text{and} \quad ATT'^{(4)}_{\theta}=\frac{17}{3}.
\end{equation}

Obviously, on the basis of the statements above, we can demonstrate that the three graphs, namely, star graph $S_{m}$, complete graph $K_{n}$ and graph $G_{1}$, all have most optimal topological structure while the left graph $G_{2}$ is not optimal. It is natural to ask what leads to such a phenomenon. In order to address this issue, the next task will aim to distinguish  topological structures among those graphs considered here. As a result, in the following sections, theorem 3 will provide a sufficient and necessary condition such that the equality of Eq.(\ref{eqa:MF-3-1-1}) may be satisfied by only star-type graphs defined in Section 2. To accomplish the development of theorem 3, we would like to first consider the trapping problem in star-type graphs.

\section{Tapping problem on star-type graph}

In this section, our main goal is to develop a detailed computation for the average trapping time $ATT^{\star}_{\theta}$ on star-type graph $G'$ based on the methods of generating functions.

\textbf{Theorem 1} \emph{For an arbitrary star-type graph $G'(V',E')$ with a trap $\theta$ allocated at that external vertex $u$, the closed-form solution to average trapping time $ATT^{\star}_{\theta}$ is given by
}
 \begin{equation}\label{eqa:MF-4-1}
ATT^{\star}_{\theta}=\frac{2|E'|}{d_{\theta}}-1
\end{equation}
\emph{where $d_{\theta}$ is the degree of that trap and in fact equals $d_{u}$.}

\emph{Proof } First of all, by demonstrations in subsections 2.1 and 2.3, we have to interpret star-type graph $G'(V',E')$ by its matrix representation as follows

$$\mathbf{A}^{\star}=\left(
  \begin{array}{cccc}
    0 & \mathbf{1}_{|V_{1}|} & \dots & \mathbf{1}_{|V_{m}|} \\
    \mathbf{1}^{\top}_{|V_{1}|} & \mathbf{A}_{|V_{1}|\times |V_{1}|} & \dots & \mathbf{0}_{|V_{1}|\times |V_{m}|}  \\
     \vdots  & \vdots  & \ddots & \vdots  \\
     \mathbf{1}^{\top}_{|V_{m}|} & \mathbf{0}^{\top}_{|V_{1}|\times |V_{m}|}  & \dots  & \mathbf{A}_{|V_{m}|\times |V_{m}|} \\
  \end{array}\right)_{|V'|\times |V'|}
  $$
in which $\mathbf{A}_{|V_{i}|\times |V_{i}|}$ is the adjacency matrix of each component $G_{i}(V_{i},E_{i})$ in star-type graph $G'(V',E')$ and symbol $\mathbf{1}_{|V_{i}|}$ denotes a $|V_{i}|$-dimensional vector whose elements are all $1$, i.e., $$\mathbf{1}_{|V_{i}|}=(\underbrace{1,1,\dots,1}_{|V_{i}|}).$$ And, the superscript $\top$ represents transpose. For our propose, the above matrix $\mathbf{A}^{\star}$ can be simplified as

$$\mathbf{A}^{\star}=\left(
      \begin{array}{cc}
        0 & \mathbf{1} \\
        \mathbf{1}^{\top} & \mathbf{A}^{\star}_{\theta} \\
      \end{array}
    \right)$$
where $\mathbf{1}=(\underbrace{1,1,\dots,1}_{\sum_{i=1}^{m}|V_{i}|})$.

Now, let us turn our attention on the development of proof of Eq.(\ref{eqa:MF-4-1}). To this end, by definition in Eq.(\ref{eqa:MF-2-4-1}), we should first determine the formula for the trapping time $TT^{\star}_{v\rightarrow\theta}$ of each vertex $v$ in star-type graph $G'(V',E')$. For convenience, each vertex $v$ will be assigned a distinct number. As an illustrative example, we may denote by $TT^{\star}_{v_{j}^{i}\rightarrow\theta}$ the trapping time of vertex $v^{i}_{j}$ in component $G_{i}(V_{i},E_{i})$. As shown in Section 3, one can make use of method from spectral graph theory to derive exact solution to the trapping time $TT^{\star}_{v_{j}^{i}\rightarrow\theta}$. Here, we intend to capture the admirable results in a manner based on generating functions. Thus, we have to introduce some notations, as follows

\begin{itemize}
\item Notation $P_{v_{j}^{i}\rightarrow\theta}(s)$ is defined as the probability that a walker first arrive at that trap $\theta$ from its original location $v_{j}^{i}$ in component $G_{i}(V_{i},E_{i})$ after $s$ steps.

Using such a description, we can integrate probability $P_{v_{j}^{i}\rightarrow\theta}(s)$ for each vertex in the following form

$$\mathbf{P}(s)=\left(P_{v_{1}^{1}\rightarrow\theta}(s), P_{v_{2}^{1}\rightarrow\theta}(s), \dots, P_{v_{|V_{1}|}^{1}\rightarrow\theta}(s), P_{v_{1}^{2}\rightarrow\theta}(s),\dots,P_{v_{|V_{m}|}^{m}\rightarrow\theta}(s)\right).$$

\item  Notation $\mathcal{P}_{v_{j}^{i}\rightarrow\theta}(z)$ is viewed as the corresponding probability generating function of quantity $P_{v_{j}^{i}\rightarrow\theta}(s)$.

Analogously, we can write $\mathcal{P}(z)$ in vector term

$$\mathcal{P}(s)=\left(\mathcal{P}_{v_{1}^{1}\rightarrow\theta}(s), \mathcal{P}_{v_{2}^{1}\rightarrow\theta}(s), \dots, \mathcal{P}_{v_{|V_{1}|}^{1}\rightarrow\theta}(s), \mathcal{P}_{v_{1}^{2}\rightarrow\theta}(s),\dots,\mathcal{P}_{v_{|V_{m}|}^{m}\rightarrow\theta}(s)\right).$$

\item According to subsection 2.1, two $\left(\sum_{i=1}^{m}|V_{i}|\right)$-dimensional vectors are introduced in terms of

  $$\mathbf{D}_{\theta}=\left(d_{v_{1}^{1}},d_{v_{2}^{1}}, \dots, d_{v_{|V_{1}|}^{1}}, d_{v_{1}^{2}}, \dots, d_{v_{|V_{m}|}^{m}}\right), \quad\mathbf{K}_{\theta}=\left(\frac{1}{d_{v_{1}^{1}}},\frac{1}{d_{v_{2}^{1}}}, \dots, \frac{1}{d_{v_{|V_{1}|}^{1}}},\frac{1}{ d_{v_{1}^{2}}}, \dots, \frac{1}{d_{v_{|V_{m}|}^{m}}}\right)$$

Note, $d_{v_{j}^{i}}$ is the degree of vertex $v_{j}^{i}$ in component $G_{i}(V_{i},E_{i})$ of star-type graph $G'(V',E')$.

\end{itemize}

After that, it is not hard using the notation mentioned above to see that all the master equations for probabilities $P_{v_{i}^{j}\rightarrow\theta}(s)$ obey the coming vector representation

 \begin{equation}\label{eqa:MF-4-2}
\mathbf{P}(s)^{\top}=\mathbf{D}_{\theta}^{-1}\mathbf{A}^{\star}_{\theta}\mathbf{P}(s-1)^{\top}+\delta_{s,1}\mathbf{K}_{\theta}^{\top},
\end{equation}
where $\delta_{s,1}$ is the Kronecker delta function in which $\delta_{s,1}=1$ when $s=1$ and $\delta_{s,1}=0$ otherwise.

Multiplying $z^{s}$ on the both-hand sides in Eq.(\ref{eqa:MF-4-2}) and using characteristic of generating functions together produces

\begin{equation}\label{eqa:MF-4-3}
\mathcal{P}(z)^{\top}=z\left(\mathbf{D}_{\theta}^{-1}\mathbf{A}^{\star}_{\theta}\mathcal{P}(z)^{\top}+\mathbf{K}_{\theta}^{\top}\right).
\end{equation}

To make further progress, taking the differential on the both sides of Eq.(\ref{eqa:MF-4-3}) yields

\begin{equation}\label{eqa:MF-4-4}
\left.\dfrac{d}{dz}\mathcal{P}(z)^{\top}\right|_{z=1}=\left.\dfrac{d}{dz}z\left(\mathbf{D}_{\theta}^{-1}\mathbf{A}^{\star}_{\theta}\mathcal{P}(z)^{\top}+\mathbf{K}_{\theta}^{\top}\right)\right|_{z=1}.
\end{equation}

It is noteworthy that we are going to take useful advantage of a typical result shown by probability generating functions themselves. More specifically, for a walker starting out from vertex $v_{j}^{i}$, the expected trapping time $TT^{\star}_{v_{j}^{i}\rightarrow\theta}$ can be solved to read

\begin{equation}\label{eqa:MF-4-5}
TT^{\star}_{v_{j}^{i}\rightarrow\theta}=\left.\dfrac{d}{dz}\mathcal{P}_{v_{j}^{i}}(z)\right|_{z=1}.
\end{equation}

Thus, Eq.(\ref{eqa:MF-4-6}) can be rearranged as

\begin{equation}\label{eqa:MF-4-6}
\mathbf{TT}^{\top}=\mathbf{1}^{\top}+\mathbf{D}_{\theta}^{-1}\mathbf{A}^{\star}_{\theta}\mathbf{TT}^{\top},
\end{equation}
here we have used
$$\mathbf{TT}=\left(TT^{\star}_{v_{1}^{1}\rightarrow\theta}, TT^{\star}_{v_{2}^{1}\rightarrow\theta}, \dots, TT^{\star}_{v_{|V_{1}|}^{1}\rightarrow\theta}, TT^{\star}_{v_{1}^{2}\rightarrow\theta},\dots,TT^{\star}_{v_{|V_{m}|}^{m}\rightarrow\theta}\right).$$

Also, we can rewrite Eq.(\ref{eqa:MF-4-6}) as

\begin{equation}\label{eqa:MF-4-7}
\mathbf{D}_{\theta}\mathbf{TT}^{\top}=\mathbf{D}_{\theta}\mathbf{1}^{\top}+\mathbf{A}^{\star}_{\theta}\mathbf{TT}^{\top}.
\end{equation}

Notice that using some simple arithmetics, Eq.(\ref{eqa:MF-4-7}) in fact indicates

\begin{equation}\label{eqa:MF-4-8}
\sum_{i=1}^{m}\sum_{j=1}^{|V_{i}|}d_{v_{j}^{i}}TT^{\star}_{v_{j}^{i}\rightarrow\theta}=\sum_{i=1}^{m}\sum_{j=1}^{|V_{i}|}d_{v_{j}^{i}}+\sum_{i=1}^{m}\sum_{j=1}^{|V_{i}|}(d_{v_{j}^{i}}-1)TT^{\star}_{v_{j}^{i}\rightarrow\theta}.
\end{equation}

Clearly, we may have

\begin{equation}\label{eqa:MF-4-9}
\sum_{i=1}^{m}\sum_{j=1}^{|V_{i}|}TT_{v_{j}^{i}\rightarrow\theta}=\sum_{i=1}^{m}\sum_{j=1}^{|V_{i}|}d_{v_{j}^{i}}.
\end{equation}

By using an obvious result $\sum_{i=1}^{m}\sum_{j=1}^{|V_{i}|}d_{v_{j}^{i}}=2|E'|-d_{u}$, we can by definition state that the exact solution to average trapping time $ATT^{\star}_{\theta}$ is

\begin{equation}\label{eqa:MF-4-10}
ATT^{\star}_{\theta}=\frac{\sum_{i=1}^{m}\sum_{j=1}^{|V_{i}|}d_{v_{j}^{i}}TT^{\star}_{v_{j}^{i}\rightarrow\theta}}{|V'|-1}=\frac{2|E'|-d_{u}}{|V'|-1}.
\end{equation}

By far, it is clear to the eye that Eq.(\ref{eqa:MF-4-10}) is the same as Eq.(\ref{eqa:MF-4-1}) after utilizing a fact $|V'|-1=d_{u}$. Theorem 1 is complete.

From the demonstration above, one can easily observe that all star-type graphs have most optimal trapping efficiency by making the equality in Eq.(\ref{eqa:MF-3-1-1}) true. This suggests that the necessity of theorem 3 (shown in detail later) is valid. Yet, in order to complete the theorem 3 successfully, we need to take many other works as shown shortly. To put this another way, the sufficiency of theorem 3 will be left to prove in the next section.

The reminding of this section is to derive the closed-form expression of another structural parameter of star-type graphs that is usually called eigentime identity \cite{Aldous-1999} or often thought of as Kemeny's constant $\overline{ATT_{\theta}}$. This parameter is in essence related to the trapping problem on graph $G(V,E)$ with a single trap $\theta$ and is expressed as

 \begin{equation}\label{eqa:MF-4-11}
\overline{ATT_{\theta}}=\sum_{v\in V}\pi_{v}TT_{v\rightarrow\theta}.
\end{equation}

It is straightforward to find that there is a noticeable relation between average trapping time $ATT_{\theta}$ and Kemeny's constant $\overline{ATT_{\theta}}$, i.e.,

  \begin{equation}\label{eqa:MF-4-12}
\overline{ATT_{\theta}}=(1-\pi_{\theta})ATT_{\theta}.
\end{equation}

Then, as an immediate consequence, we can obtain upon Eqs.(\ref{eqa:MF-4-1}) and (\ref{eqa:MF-4-12}) the following theorem with omitting analytical proof.

\textbf{Theorem 2} \emph{For an arbitrary Star-type graph $G'(V',E')$, when the trap $\theta$ is allocated at that external vertex $u$, the closed form of Kemeny's constant $\overline{ATT'_{\theta}}$ can be precisely given by
}
\begin{equation}\label{eqa:MF-4-13}
\overline{ATT'_{\theta}}=\frac{2|E'|}{|V'|-1}+\frac{|V'|-1}{2|E'|}-2.
\end{equation}

In addition, by definition of average degree $\langle k\rangle$ of graph $G(V,E)$, namely, $\langle k\rangle=2|E|/|V|$, the value for $\overline{ATT'_{\theta}}$ in Eq.(\ref{eqa:MF-4-13}) can be approximately equal to

\begin{equation}\label{eqa:MF-4-14}
\overline{ATT'_{\theta}}\approx \langle k'\rangle + \frac{1}{\langle k'\rangle}-2.
\end{equation}

Armed with Eqs.(\ref{eqa:MF-4-1}) and (\ref{eqa:MF-4-14}), we can certainly stress that when allocating the trap $\theta$ on the external vertex $u$ in star-type graph $G'(V',E')$ in question, the values for two quantities, average trapping time and Kemeny's constant, may be asymptotically determined by the average degree $\langle k'\rangle$.

In a word, the analytical solution to average trapping time for star-type graphs considered has been completely derived. Additionally, with these results above, one will be able to find out some potential applications both theoretical and practical as will be shown shortly.

\section{Theoretical applications}

From the theoretical point of view, the lights shed by developing Section 4 can help us to study the trapping problem in some graphs of significant interest and then exactly determine the corresponding solutions to average trapping time. Here, what we are interested in is an ensemble of graphs $\mathcal{G}^{n}$ which are obtained from star-type graphs $G'$ by using $n$th-order subdivision operation defined in subsection 2.3. Before beginning our discussions, the concrete construction of graphs $\mathcal{G}^{n}$ can be found in the following form

\begin{itemize}
\item Graphs $\mathcal{G}^{n}$ are generated from graphs $G'$ by applying $n$th-order subdivision operation to each edge $e_{v^{i}_{j}v^{i}_{l}}$ in component $G_{i}$. Those number $n$ newly inserted vertices to edge $e_{v^{i}_{j}v^{i}_{l}}$ may be labelled $v^{i}_{jl}(t)$ $(t=1,2,...,n)$ for our purpose. And then, we may group all vertices in component $G^{n}_{i}$ of graph $\mathcal{G}^{n}$ into two sets, one set $V_{i}$ consisting of vertices $v^{i}_{j}$ and all other vertices $v^{i}_{j}(t)$ contained in set $V_{i}(n)$. In other words, the vertex set $V_{i}^{n}$ of graph $G^{n}_{i}$ is constituted by both sets $V_{i}$ and $V_{i}(n)$, i.e., $V_{i}^{n}=V_{i}\cup V_{i}(n)$.

Apparently, the values for vertex number $|\mathcal{V}^{n}|$ and edge number $|\mathcal{E}^{n}|$ of the resulting graphs $\mathcal{G}^{n}$ can be easily obtained based on Eq.(\ref{eqa:MF-2-3-2}), as below

 \begin{equation}\label{eqa:MF-5-1}
|\mathcal{V}^{n}|=|V'|+n\sum_{i=1}^{m}|E_{i}|, \qquad |\mathcal{E}^{n}|=d_{u}+(n+1)\sum_{i=1}^{m}|E_{i}|.
\end{equation}

\end{itemize}

Next, from Eqs(\ref{eqa:MF-3-1-1}) and (\ref{eqa:MF-5-1}), lemma 2 can be read as follows

\textbf{Lemma 2}  \emph{For an arbitrary graph $\mathcal{G}^{1}$ with a trap $\theta$ allocated on external vertex $u$, the theoretical lower bound for $\mathcal{ATT}^{1}_{\theta}$ is given by}

 \begin{equation}\label{eqa:MF-5-2}
\mathcal{ATT}^{1}_{\theta}\geq\frac{4\sum_{i=1}^{m}|E_{i}|}{d_{\theta}}+1
\end{equation}
\emph{where $d_{\theta}$ is in fact equal to the degree of vertex $u$ and $|E_{i}|$ is the total number of edges of component $G_{i}$ in star-type graph $G'$.}

On the other hand, the upper bound for average trapping time $\mathcal{ATT}^{1}_{\theta}$ is shown in proposition 1.

\textbf{Proposition 1}\emph{ For an arbitrary graph $\mathcal{G}^{1}$ with a trap $\theta$ allocated on external vertex $u$, the upper bound for $\mathcal{ATT}^{1}_{\theta}$ is written as}

 \begin{equation}\label{eqa:MF-5-3}
\mathcal{ATT}^{1}_{\theta}<\frac{4(\sum_{i=1}^{m}|E_{i}|)^{2}+d_{\theta}\sum_{i=1}^{m}|E_{i}|+4\sum_{i=1}^{m}|E_{i}|}{d_{\theta}+\sum_{i=1}^{m}|E_{i}|}+1
\end{equation}
\emph{where symbols $|E_{i}|$ and $d_{\theta}$ both have the same meanings as shown in lemma 2.}

\emph{Proof} By analogy with the proof of theorem 1, we first need to take some helpful notations in order to consolidate proposition 1. It is worth marking that in the following, we only provide simple descriptions about notations used later for convenience and reader is encouraged to refer to the development of theorem 1 for more detail.

\begin{itemize}
\item The adjacency matrix of graph $\mathcal{G}^{1}$ is expressed as

$$\mathbf{A}^{1}=\left(
  \begin{array}{ccccccc}
    0 & \mathbf{1}_{|V_{1}|} & \mathbf{0}_{|E_{1}|} &\dots & \mathbf{1}_{|V_{m}|} & \mathbf{0}_{|E_{m}|} \\
    \mathbf{1}^{\top}_{|V_{1}|} & \mathbf{0}_{|V_{1}|\times |V_{1}|} & \mathbf{B}_{|V_{1}|\times |E_{1}|} & \dots & \mathbf{0}_{|V_{1}|\times |V_{m}|}  & \mathbf{0}_{|V_{1}|\times |E_{m}|} \\
     \mathbf{0}^{\top}_{|E_{1}|} & \mathbf{B}^{\top}_{|V_{1}|\times |E_{1}|}  & \mathbf{0}_{|E_{1}|\times |E_{1}|}  & \dots & \mathbf{0}_{|E_{1}|\times |V_{m}|}  & \mathbf{0}_{|E_{1}|\times |E_{m}|} \\
     \vdots  & \vdots  & \vdots & \ddots  & \vdots  & \vdots \\
     \mathbf{1}^{\top}_{|V_{m}|} & \mathbf{0}^{\top}_{|V_{1}|\times |V_{m}|} & \mathbf{0}^{\top}_{|E_{1}|\times |V_{m}|} & \dots  & \mathbf{0}_{|V_{m}|\times |V_{m}|}& \mathbf{B}_{|V_{m}|\times |E_{m}|}\\
     \mathbf{0}^{\top}_{|E_{m}|}& \mathbf{0}^{\top}_{|V_{1}|\times |E_{m}|} & \mathbf{0}^{\top}_{|E_{1}|\times |E_{m}|} & \dots & \mathbf{B}^{\top}_{|V_{m}|\times |E_{m}|}& \mathbf{0}_{|E_{m}|\times |E_{m}|}\\
  \end{array}\right)_{|\mathcal{V}^{1}|\times |\mathcal{V}^{1}|}
  $$

\item Notation $P^{1}_{v_{j}^{i}\rightarrow\theta}(s)$ is thought of as the probability that a walker starting out from its initial location $v_{j}^{i}$ in vertex set $V_{i}$ of component $G^{1}_{i}(V^{1}_{i},E^{1}_{i})$ first hit that trap $\theta$ after $s$ steps. Similarly, notation $Q^{1}_{v_{jl}^{i}(1)\rightarrow\theta}(s)$ is defined as the probability for a walker to first hop on the trap $\theta$ from its initial vertex $v_{jl}^{i}(1)$ in component $G^{1}_{i}(V^{1}_{i},E^{1}_{i})$ in $s$ steps. Thus, we can have two vectors

$$\mathbf{P}^{1}(s)=\left(P^{1}_{v_{j}^{i}\rightarrow\theta}(s)\right),\qquad \mathbf{Q}(s)=\left(Q_{v_{jl}^{i}(1)\rightarrow\theta}(s)\right).$$

\item  Notations $\mathcal{P}^{1}_{v_{j}^{i}\rightarrow\theta}(z)$ and $\mathcal{Q}_{v_{jl}^{i}(1)\rightarrow\theta}(z)$ are regarded as the corresponding probability generating functions of quantities $P^{1}_{v_{j}^{i}\rightarrow\theta}(s)$ and $Q_{v_{jl}^{i}(1)\rightarrow\theta}(s)$, respectively. As above, two vectors corresponding to probability generating functions introduced can be read

$$\mathcal{P}^{1}(s)=\left(\mathcal{P}^{1}_{v_{j}^{i}\rightarrow\theta}(s)\right),\qquad \mathcal{Q}(s)=\left(\mathcal{Q}_{v_{jl}^{i}(1)\rightarrow\theta}(s)\right).$$

\item Following the preceding descriptions, we should take two $\left(|\mathcal{V}^{1}|-1\right)$-dimensional vectors in terms of

  $$\mathbf{D}^{1}_{\theta}=\left(d_{v_{1}^{1}},d_{v_{2}^{1}}, \dots, d_{v_{|V_{1}|}^{1}}, d_{v_{1}^{2}}, \dots, d_{v_{|V_{m}|}^{m}}, \underbrace{2,2,...,2}_{\sum_{i=1}^{m}|E_{i}|}\right)$$
  and $$ \quad\mathbf{K}^{1}_{\theta}=\left(\frac{1}{d_{v_{1}^{1}}},\frac{1}{d_{v_{2}^{1}}}, \dots, \frac{1}{d_{v_{|V_{1}|}^{1}}},\frac{1}{ d_{v_{1}^{2}}}, \dots, \frac{1}{d_{v_{|V_{m}|}^{m}}}, \underbrace{\frac{1}{2},\frac{1}{2},...,\frac{1}{2}}_{\sum_{i=1}^{m}|E_{i}|}\right).$$

\item Final, we denote by $TT^{1}_{v_{j}^{i}\rightarrow\theta}$ ($TT^{1}_{v_{jl}^{i}(1)\rightarrow\theta}$) the average trapping time for a walker who originally starts out from its location $v_{j}^{i}$ ($v_{jl}^{i}(1)$).

\end{itemize}

According to similar computations as shown in the process of proving theorem 1, we may come to the next both equations

\begin{subequations}
\label{eq:whole}
\begin{eqnarray}
\sum_{i=1}^{m}\sum_{j=1}^{|V^{1}_{i}|}(d_{v_{j}^{i}}+1)TT^{1}_{v_{j}^{i}\rightarrow\theta}=
\sum_{i=1}^{m}\sum_{j=1}^{|V^{1}_{i}|}(d_{v_{j}^{i}}+1)+\sum_{i=1}^{m}\sum_{j=1}^{|V^{1}_{i}|}d_{v_{j}^{i}}TT^{1}_{v_{j}^{i}\rightarrow\theta}+
2\sum_{i=1}^{m}|E_{i}|,\label{subeq:MF-5-4-1}
\end{eqnarray}
\begin{equation}
2\sum_{i=1}^{m}\sum_{v_{jl}^{i}(1)\in V_{i}(1)}TT^{1}_{v_{jl}^{i}(1)\rightarrow\theta}=2\sum_{i=1}^{m}|E_{i}|+\sum_{i=1}^{m}\sum_{j=1}^{|V^{1}_{i}|}d_{v_{j}^{i}}TT^{1}_{v_{j}^{i}\rightarrow\theta}.\label{subeq:MF-5-4-2}
\end{equation}
\end{subequations}

To make further progress, Eq.(\ref{subeq:MF-5-4-1}) can be reorganized to obtain

 \begin{equation}\label{eqa:MF-5-5}
\sum_{i=1}^{m}\sum_{j=1}^{|V^{1}_{i}|}TT^{1}_{v_{j}^{i}\rightarrow\theta}=4\sum_{i=1}^{m}|E_{i}|+\sum_{i=1}^{m}|V^{1}_{i}|
\end{equation}
here we have used an obvious result $\sum_{i=1}^{m}\sum_{j=1}^{|V^{1}_{i}|}d_{v_{j}^{i}}=2\sum_{i=1}^{m}|E_{i}|$.

By definition in Eq.(\ref{eqa:MF-2-4-1}), it is not difficult to find the precise formula for average trapping time $\mathcal{ATT}^{1}_{\theta}$ to be

 \begin{equation}\label{eqa:MF-5-6}
\mathcal{ATT}^{1}_{\theta}=\frac{5\sum_{i=1}^{m}|E_{i}|+\sum_{i=1}^{m}|V^{1}_{i}|+\frac{1}{2}\sum_{i=1}^{m}\sum_{j=1}^{|V^{1}_{i}|}d_{v_{j}^{i}}TT^{1}_{v_{j}^{i}\rightarrow\theta}}{\sum_{i=1}^{m}(|E_{i}|+|V^{1}_{i}|)}.
\end{equation}

In order to finally come to the presentation as observed in Eq.(\ref{eqa:MF-5-3}), we have to take the well-known Holder inequality in discrete form, as follows

 \begin{equation}\label{eqa:MF-5-7}
\left(\sum_{i}^{n}a_{i}^{p}\right)^{\frac{1}{p}}\left(\sum_{i}^{n}b_{i}^{q}\right)^{\frac{1}{q}}\geq\sum_{i=1}^{n}a_{i}b_{i}.
\end{equation}

Then, we obtain

 \begin{equation}\label{eqa:MF-5-8}
\sum_{i=1}^{m}\sum_{j=1}^{|V^{1}_{i}|}d_{v_{j}^{i}}TT^{1}_{v_{j}^{i}\rightarrow\theta}<\left(2\sum_{i=1}^{m}|E_{i}|\right)\left(4\sum_{i=1}^{m}|E_{i}|+\sum_{i=1}^{m}|V^{1}_{i}|\right).
\end{equation}

Plugging both Eq.(\ref{eqa:MF-5-8}) and result $d_{\theta}=\sum_{i=1}^{m}|V^{1}_{i}|$ into Eq.(\ref{eqa:MF-5-6}) leads to the same result as said in Eq.(\ref{eqa:MF-5-3}). This suggests that proposition 1 is true.

More interestingly, if we only consider summation of the trapping times over all vertices $v^{i}_{j}$ in vertex sets $V^{1}_{i}$, then the average value $ATT^{1}_{\theta}$ can be calculated by Eq.(\ref{eqa:MF-5-5}) to obtain

 \begin{equation}\label{eqa:MF-5-9}
ATT^{1}_{\theta}=\frac{\sum_{i=1}^{m}\sum_{j=1}^{|V^{1}_{i}|}TT^{1}_{v_{j}^{i}\rightarrow\theta}}{\sum_{i=1}^{m}|V^{1}_{i}|}=\frac{4\sum_{i=1}^{m}|E_{i}|}{\sum_{i=1}^{m}|V^{1}_{i}|}+1.
\end{equation}

By adopting a trivial fact $\sum_{i=1}^{m}|V^{1}_{i}|=d_{\theta}$, the term of the right-hand side in Eq.(\ref{eqa:MF-5-9}) is completely equivalent to that of the right-hand side in previous Eq.(\ref{eqa:MF-5-2}). This implies that the value for $ATT^{1}_{\theta}$ has made the equality in Eq.(\ref{eqa:MF-5-2}) true. In addition, with the help of an apparent fact shown in the next form

 \begin{equation}\label{eqa:MF-5-10}
\begin{aligned}\left(\sum_{i=1}^{m}|V^{1}_{i}|\right)\left(\sum_{i=1}^{m}\sum_{j=1}^{|V^{1}_{i}|}d_{v_{j}^{i}}TT^{1}_{v_{j}^{i}\rightarrow\theta}\right)&\geq\left(\sum_{i=1}^{m}\sum_{j=1}^{|V^{1}_{i}|}d_{v_{j}^{i}}\right)\left(\sum_{i=1}^{m}\sum_{j=1}^{|V^{1}_{i}|}TT^{1}_{v_{j}^{i}\rightarrow\theta}\right)\\
&>8\left(\sum_{i=1}^{m}|E_{i}|\right)^{2},
\end{aligned}
\end{equation}
we can have

 \begin{equation}\label{eqa:MF-5-11}
\frac{\sum_{i=1}^{m}\sum_{j=1}^{|V^{1}_{i}|}d_{v_{j}^{i}}TT^{1}_{v_{j}^{i}\rightarrow\theta}}{\sum_{i=1}^{m}|E_{i}|}>
\frac{\sum_{i=1}^{m}\sum_{j=1}^{|V^{1}_{i}|}TT^{1}_{v_{j}^{i}\rightarrow\theta}}{\sum_{i=1}^{m}|V^{1}_{i}|}
\end{equation}
and then find

 \begin{equation}\label{eqa:MF-5-12}
\mathcal{ATT}^{1}_{\theta}=\frac{\sum_{i=1}^{m}\sum_{j=1}^{|V^{1}_{i}|}(d_{v_{j}^{i}}+1)TT^{1}_{v_{j}^{i}\rightarrow\theta}}{\sum_{i=1}^{m}|V^{1}_{i}|+\sum_{i=1}^{m}|E_{i}|}>
\frac{\sum_{i=1}^{m}\sum_{j=1}^{|V^{1}_{i}|}TT^{1}_{v_{j}^{i}\rightarrow\theta}}{\sum_{i=1}^{m}|V^{1}_{i}|}.
\end{equation}
This means that graph $\mathcal{G}^{1}$ is not most optimal in such a situation considered here.

Now, we have the ability to demonstrate theorem 3.

\textbf{Theorem 3} \emph{For a given graph $G$ with a trap $\theta$ allocated at some vertex $u$, the average trapping time $ATT_{\theta}$ can satisfy the equality in Eq.(\ref{eqa:MF-3-1-1}) if and only if the underlying structure is star-type.
}
\emph{Proof} This may be completed based on theorem 1 and proposition 1 and thus we omit it here.

As a case study, let each component $G_{i}$ in graph $G'$ be a $\mathbf{d}$-regular graph\footnote[1]{A graph $G$ is considered $\mathbf{d}$-regular if its each vertex has degree $\mathbf{d}$.} and then for the end graph $\mathcal{G}^{1}(\mathbf{d})$ constructed upon graph $G'$, we have the next corollary 1.

\textbf{Corollary 1} \emph{For the graph $\mathcal{G}^{1}(\mathbf{d})$ mentioned above, when a trap $\theta$ allocated on external vertex $u$, the closed-form solution to average trapping time $\mathcal{ATT}^{1}_{\theta}(\mathbf{d})$ follows}

 \begin{equation}\label{eqa:MF-5-13}
\mathcal{ATT}^{1}_{\theta}(\mathbf{d})=\frac{(5+2\mathbf{d})\sum_{i=1}^{m}|E_{i}|+(1+\frac{\mathbf{d}}{2})d_{\theta}}{d_{\theta}+\sum_{i=1}^{m}|E_{i}|}.
\end{equation}

More generally, from the demonstration in proposition 1, one can without more effort find out the following corollary.

\textbf{Corollary 2} \emph{For an arbitrary graph $\mathcal{G}^{n}$ with a trap $\theta$ allocated on external vertex $u$, the upper bound for $\mathcal{ATT}^{n}_{\theta}$ obeys the next equations.}

\emph{As $n$ is even,}

 \begin{equation}\label{eqa:MF-5-14}
\mathcal{ATT}^{n}_{\theta}<\frac{n(n+1)\left(\sum_{i=1}^{m}|E_{i}|\right)^{2}+\left(\frac{nd_{\theta}}{2}+2(n+1)+4\sum_{i=1}^{\frac{n}{2}}i^{2}\right)\left(\sum_{i=1}^{m}|E_{i}|\right)+d_{\theta}}{|\mathcal{V}^{n}|-1},
\end{equation}
\emph{otherwise,
}
 \begin{equation}\label{eqa:MF-5-15}
\mathcal{ATT}^{n}_{\theta}<\frac{n(n+1)\left(\sum_{i=1}^{m}|E_{i}|\right)^{2}+\left[\frac{nd_{\theta}}{2}+2(n+1)+2\left(\sum_{i=1}^{\lfloor\frac{n}{2}\rfloor}i(2i+1)\right)+\sum_{i=0}^{\lfloor\frac{n}{2}\rfloor}(2i+1)\right]\left(\sum_{i=1}^{m}|E_{i}|\right)+d_{\theta}}{|\mathcal{V}^{n}|-1}.
\end{equation}
Here symbols $|E_{i}|$ and $d_{\theta}$ both have the same meanings as shown above.

\section{Practical applications}

Here, with the help of the results above, our goal is to discuss some more practical applications on real-world networks, including scale-free ones.

It is well known that the degree is the most fundamental parameter for describing the level of vertex's importance in a graph. In general, the greater the degree of vertex $v$ in a given graph $G$, the more important vertex $v$ is. Clearly, the demonstration is, to some extent, consistent with our observation with respect to Eq.(\ref{eqa:MF-3-1-1}). Specifically, in a graph $G$ under consideration, a vertex with greater degree must have a more smaller theoretical lower bound for its corresponding quantity $ATT$. On the other hand, it is not necessary for two vertices with the same degree to exhibit identical number value for $ATT$. This reveals that the degree value is not a more available topological parameter for adequately evaluating vertex's importance, a trivial fact that has been reported in a large number of previous papers in the last. As a remedy, many other complementary measures have been made in the literature, including connectivity \cite{Bondy-2008}. Consider a graph $G$, a vertex $v$ in $G$ plays a more significant role if the deletion of vertex $v$ makes graph $G$ disconnected. On the one hand, as shown above, when star-type graph $G'$ consists only of a component, the deletion of an arbitrary vertex can not disconnect original graph. On the other hand, the external vertex $u$ in star-type graph $G'$ above can be considered more important than all other vertices based on both indices, degree and average trapping time. Therefore, this strongly implies that there is a long way to go for better quantifying vertex itself importance in a given graph \cite{Linyuan-2016,Vazquez-Araujo-2018}.

Following the concept of connectivity, there are a great variety of control schemes designed for real-life networks \cite{Albert-2000}-\cite{Shang-2019} in order to keep their robustness. The empirical observations in the context of complex networks show that the scale-free feature can be widely found in most real-life networks, where the degree distribution $P(k)$ follows

 \begin{equation}\label{eqa:MF-6-1}
P(k)\sim k^{-\gamma}.
\end{equation}
More generally, the exponent $\gamma$ in Eq.(\ref{eqa:MF-6-1}) is proven to fall into a range $2<\gamma<3$ in most situations. As an immediate consequence, we can have two obvious facts in the networks $G(V,E)$ of such type in the following form

$$|E|=O(|V|), \qquad k_{max}=O\left(|V|^{\frac{1}{1-\gamma}}\right),$$
and thus derive an approximate formula for $ATT_{\theta}$ of this kind of networks when allocating the trap $\theta$ on the greatest degree vertex, as follows

\begin{equation}\label{eqa:MF-6-2}
ATT_{\theta}\geq|V|^{1-\frac{1}{1-\gamma}},
\end{equation}
meaning which the scaling of $ATT_{\theta}$ grows sublinearly with the vertex number $|V|$.

Fortunately, various scale-free networks \cite{Agliari-2009,Zhang-2011}, both stochastic and deterministic, turn out to not be optimal because they fail to hold the equality of Eq.(\ref{eqa:MF-6-2}). In fact, it is an obvious deduction from theorem 3. That is to say, the underlying structures corresponding to such networks can not be star type. In order to produce scale-free networks with optimal trapping efficiency, a more effective measure is to connect more vertices to a designated object. Motivated by this, a principle framework for succeeding in addressing this problem has been proposed in our recent work \cite{Ma-2020-PRE}.

From the discussions above, we can easily see that it is not likely to completely control scale-free networks with degree exponent $2<\gamma<3$ via placing a single trap on an arbitrary vertex. Therefore, we should assign some traps on vertices in the networks of this type to make trapping efficiency more optimal. Fortunately, the characteristic of these such networks imply that there a small fraction of vertices possessing a large portion of connections. So, those vertices should be occupied by  traps for reaching our desirable result. It is worth noting that many similar demonstrations to our conclusion have been developed from the other theoretical points of view, such as the concept of dominating set \cite{Bondy-2008,Chen-2016,Chalermsook-2017}. Armed with both the statements and theorem 3, we come to the last corollary in the paper, as below

\textbf{Corollary 3} \emph{For a given graph $G$, a set $S$ of vertices compose a dominating set if and only if the average trapping time $ATT$ can make corresponding theoretical lower bound true when each vertex in set $S$ is captured by a trap.}

\section{Conclusion and future work}

 To conclude, we study the trapping problem on graph and obtain some interesting results related to a topological parameter, defined average trapping time $ATT$, quantifying trapping efficiency on graph under consideration. By using techniques from spectral graph theory, we derive exact formula for the theoretical lower bound of quantity $ATT$ on a graph where there is a single trap. Several specific graphs are selected to show that they may have most trapping efficiency via completely achieving the corresponding lower bound for $ATT$, suggesting which the theoretical lower bound presented here is sharp. However, there are example graphs proven to not be most optimal with regard to quantity $ATT$.

To look for all graphs satisfying the theoretical lower bound of parameters $ATT$ as shown in Eq.(\ref{eqa:MF-3-1-1}), we introduce a family of graphs that is called star-type in this paper. Then, using the method of probability generating functions, we find the graphs of such type to possess most optimal trapping efficiency. As an immediate consequence, we analytically obtain the closed-form solution to another parameter, called Kemeny's constant, associated with the trapping problem when the trap is allocated on the external vertex on star-type graph. To further process, we prove that a graph is most optimal with respect to average trapping time if and only if the underlying structure is star type finally.

In addition, with the results obtained, we also consider the trapping problem on many other graphs of great interest, and analytically calculate some approximate formulas for their own quantities $ATT$. Among which, as shown in corollary 1, the equality holds for example graphs in question. From the point of view of practical application, we state that the value for $ATT$ corresponding to each vertex $v$ in a given graph can be thought of as a measurable index for describing the importance of vertex $v$. In some cases, we observe that such an index may outperform other previous parameters for evaluating vertex's importance, such as vertex degree. Last but not least, we investigate the trapping problem on scale-free networks with power exponent $2<\gamma<3$ and find that the scaling of $ATT$ may grows sublinearly with the vertex number in the thermodynamic limit. In terms of this case, we provide some available control schemes for strengthening the robustness of networks of this kind. For example, a network must display most optimal trapping efficiency only if each vertex in any dominating set is occupied by a trap as reported in corollary 3.

We, however, would like to stress that our results only provide theoretical analysis about the lower bound for quantity $ATT$. While the theoretical lower bound for $ATT$ may be satisfied by some vertex in a graph $G$, the analytic values for $ATT$ of all other vertices can not be precisely calculated using our methods here. Hence, it is more demandable to design effective algorithms for addressing this issue. One of most important reasons for this is that such an index can be used to estimate the importance of vertex itself. There in fact are a large number of real-life applications developed based on vertex's importance, for instance, graph mining and recommendation system \cite{Linyuan-2016,Prado-2013,Chen-2017}.

\section*{ACKNOWLEDGMENT}

We would like to thank Bing Yao for some useful discussions. The research was supported by the National Key Research and Development Plan under grant 2017YFB1200704 and the National Natural Science Foundation of China under grant No. 61662066.


{\footnotesize
\begin{thebibliography}{9}

\setlength{\parskip}{0pt}

\bibitem{Dongari-2011} N. Dongari, Y. Zhang, and J.M. Reese. Molecular free path distribution in rarefied gases. J. Phys. D Appl. Phys. 44(12):125502, 2011.

\bibitem{Dobson-2003} C.M. Dobson. Protein folding and misfolding. Nature. 426:884-890, 2003.

\bibitem{Guille-2012} A. Guille and H. Hacid. In Proceedings of the 21st International Conference Companion on World Wide Web (WWW), pages 1145-1152, 2012.

\bibitem{Gonzalez-2008} M.C. Gonzalez, C.A. Hidalgo, and A.-L. Barab\'{a}si. Understanding individual human mobility patterns. Nature. 453:779-782, 2008.

\bibitem{Wang-2010} J. Wang, Q.Y. Wang, and J.G. Shao. Fluctuations of stock price model by statistical physics systems. Math. Comput. Model. 51(5):431-440, 2010.

\bibitem{Dorigo-2005} M. Dorigo and C. Blum. Ant colony optimization theory: a survey. Theor. Comput. Sci. 344:243-278, 2005.

\bibitem{Li-2016} J.Z. Li, J. Zhu, and B. Zhang. Discriminative deep random walk for network classification. In pressing of the 54th Annual Meeting of the Association for Computational Linguistics (ACL), pages 1004-1013, 2016.

\bibitem{Perkins-2014} T.J. Perkins, E. Foxall, L. Glass and R. Edwards. A scaling law for random walks on networks. Nat. Commun. 5:5121, 2014.

\bibitem{Cohen-2016} M.B. Cohen, J. Kelner, J. Peebles, R.Peng, A. Sidford, and A. Vladu. Faster algorithms for computing the stationary distribution, simulating random walks, and more. In Proceedings of IEEE 57th Annual Symposium on Foundations of Computer Science (FOCS), pages 583-592, 2016.

\bibitem{Rieger-2004}  J.D. Noh and H. Rieger. Random walks on complex networks. Phys. Rev. Lett. 92(11):118701, 2004.

\bibitem{Bertasius-2017} G. Bertasius, L. Torresani, S.X. Yu, and J.H. Shi. Convolutional Random Walk Networks for Semantic Image Segmentation. In pressing of the IEEE Conference on Computer Vision and Pattern Recognition (CVPR), pages 858-866, 2017.

\bibitem{Clark-2019} A. Clark, B. Alomair, L. Bushnell, and R. Poovendran. On the structure and computation of random walk times in finite graphs. IEEE Trans. Automat. Contr. 64(11):4470-4483, 2019.

\bibitem{Volchenkov-2011}  D. Volchenkov. Random walks and flights over connected graphs and complex networks. Commun. Nonlinear Sci. Numer. Simul. 16(1):21-55, 2011.

\bibitem{Condamin-2007} S. Condamin, O. B\'{e}nichou , V. Tejedor, R. Voituriez, and J. Klafter. First-passage times in complex scale-invariant media. Nature.  450:77-80, 2007.

\bibitem{Czumaj-2011} A. Czumaj, M. Monemizadeh, K. Onak, and C. Sohler. Planar graphs: Random walks and bipartiteness testing. In Proceedings of IEEE 52th Annual Symposium on Foundations of Computer Science (FOCS), pages 423-432, 2011.

\bibitem{Guerin-2016} T. Gu\'{e}rin, N. Levernier, O. B\'{e}nichou, and R. Voituriez. Mean first-passage times of non-Markovian random walkers in confinement. Nature. 534:356-359, 2016.

\bibitem{Montroll-1969}  E.W. Montroll. Random Walks on Lattices. III. Calculation of first-passage times with application to exciton trapping on photosynthetic units. J. Math. Phys. 10:753, 1969.

\bibitem{Garza-Lopez-2005} R.A. Garza-L\'{o}pez and J.J. Kozak. Invariance relations for random walks on square-planar lattices. Chem. Phys. Lett. 406(1):38-43, 2005.

\bibitem{Bentz-2010} J.L. Bentz, J.W. Turner, and J.J. Kozak. Analytic expression for the mean time to absorption for a random walker on the Sierpinski gasket. II. The eigenvalue spectrum. Phys. Rev. E. 82(1):011137, 2010.

\bibitem{Agliari-2008} E. Agliari. Exact mean first-passage time on the T-graph. Phys. Rev. E. 77(1):011128, 2008.

\bibitem{Wu-2012} B. Wu, Y. Lin, Z.Z. Zhang, and G.R. Chen. Trapping in dendrimers and regular hyperbranched polymers. J. Chem. Phys. 137:044903, 2012.

\bibitem{Gurtovenko-2005} A.A. Gurtovenko and A. Blumen. Generalized gaussian structures: Models for polymer systems with complextopologies.  Adv. Polym. Sci. 182:171, 2005.

\bibitem{Redner-2001} S. Redner. A Guide to First-Passage Processes. Cambridge university press. 2001.

\bibitem{Kittas-2008} A. Kittas, S. Carmi, S. Havlin, and P. Argyrakis. Trapping in complex networks. EPL. 84(8):4008, 2008.

\bibitem{Albert-1999-1} A.-L. Barab\'{a}si and R. Albert. Emergence of Scaling in Random Networks. Science. 5439:509-512,  1999.

\bibitem{Watts-1998} D.J. Watts and S.H. Strogatz. Collective dynamics of small-world networks. Nature. 393:440-442, 1998.

\bibitem{Xing-2017} C.M. Xing, Y.G. Zhang, J. Ma, L. Yang, and L. Guo. Exact solutions for average trapping time of random walks on weighted scale-free networks. Fractals. 5(2):1750013, 2017.

\bibitem{Zhang-2013} Z.Z. Zhang, T. Shan, and G.R. Chen. Random walks on weighted networks. Phys. Rev. E. 87(1):012112, 2013.

\bibitem{Couto-2016} R.de.S. Couto, S. Secci, M.E.M. Campista, and L.H.M.K. Costa. Reliability and Survivability Analysis of Data Center Network Topologies. J. Netw. Syst. Manag. 24:346-392, 2016.

\bibitem{Aldous-1999} D. Aldous and J. Fill. Reversible Markov chains and random
walks on graphs. http://www.stat.berkeley.edu/~aldous/RWG/Chap2.pdf. 1999.

\bibitem{Burnashev-2012} M.V. Burnashev and A. Tchamkerten. Estimating a random walk first-passage time from noisy or delayed observations. IEEE Trans. Inf. Theory. 58(7):4230-4243, 2012.

\bibitem{Rhee-2011} I. Rhee, M. Shin, S. Hong, K. Lee, S. J. Kim, and S. Chong. On theLevy-walk nature of human mobility. IEEE/ACM Trans. Netw. 19(3):630-643, 2011.

\bibitem{Bondy-2008} J.A. Bondy and U.S.R. Murty. Graph Theory. Springer press. 2008.

\bibitem{Linyuan-2016} L.Y. L{\"u}, D.B. Chen, X.L. Ren, Q.M. Zhang, Y.-C. Zhang, and T. Zhou. Vital nodes identification in complex networks. Physics Reports. 650:1-63, 2016.

\bibitem{Vazquez-Araujo-2018} F. Vazquez-Araujo, A. Dapena, M.J. Souto-Salorio, and P.M. Castro. Calculation of the Connected Dominating Set Considering Vertex Importance Metrics. Entropy. 20(2):87, 2018.

\bibitem{Albert-2000} R. Albert, H. Jeong, and A.-L. Barab\'{a}si. Error and attack tolerance of complex networks. Nature. 406:378-382, 2000.

\bibitem{Liu-2001} Y.-Y. Liu, J.-J. Slotine, and A.-L. Barab\'{a}si. Controllability of complex networks. Nature. 473:167-173, 2011.

\bibitem{Shang-2019} Y. L. Shang. Subgraph robustness of complex networks under attacks. IEEE Transactions on Systems, Man, and Cybernetics: Systems. 49(4):821-832, 2019.

\bibitem{Agliari-2009} E. Agliari and R. Burioni. Random walks on deterministic scale-free networks: Exact results. Phys. Rev. E. 80(3):031125, 2009.

\bibitem{Zhang-2011} Z.Z. Zhang, A. Julaiti, B.Y. Hou, H.J. Zhang, and G.R. Chen. Mean first-passage time for random walks on undirected networks. Eur. Phys. J. B. 84:691-697, 2011.

\bibitem{Ma-2020-PRE} F. Ma, X.M. Wang, and P. Wang. Scale-free networks with invariable diameter and density feature: Counterexamples. Phys. Rev. E. 101(2):022315, 2020.

\bibitem{Chen-2016} Y.J. Chen and B.K. Lin. The Constant Inapproximability of the Parameterized Dominating Set Problem. In Proceedings of IEEE 57th Annual Symposium on Foundations of Computer Science (FOCS), pages 505-514, 2016.

\bibitem{Chalermsook-2017} P. Chalermsook, M. Cygan, G. Kortsarz, B. Laekhanukit, P. Manurangsi, D. Nanongkai, and L. Trevisan. From Gap-ETH to FPT-inapproximability: Clique, dominating set, and more. In Proceedings of IEEE 58th Annual Symposium on Foundations of Computer Science (FOCS), pages 743-754, 2017.

\bibitem{Prado-2013} A. Prado, M. Plantevit, C. Robardet, and J.-F. Boulicaut. Mining graph topological patterns: Finding covariations among vertex descriptors. IEEE Trans. Knowl. Data. En. 25(9):2090-2104, 2013.

\bibitem{Chen-2017} L.J. Chen, Z.K. Zhang, J.H. Liu, J. Gao, and T. Zhou. A vertex similarity index for better personalized recommendation. Physica A. 466:607-615, 2017.







\end{thebibliography}
}
\end{document}